\documentclass[journal]{IEEEtran}
\usepackage{amsmath,amsfonts}
\usepackage{algorithmic}
\usepackage{algorithm}
\usepackage{array}
\usepackage[caption=false,font=normalsize,labelfont=sf,textfont=sf]{subfig}
\usepackage{textcomp}
\usepackage{stfloats}
\usepackage{url}
\usepackage{verbatim}
\usepackage{graphicx}
\usepackage{cite}
\usepackage{xcolor}
\begin{document}
\title{Optical calibration of \\ holographic acoustic tweezers}
\author{Sonia Marrara, David Bronte Ciriza, Alessandro Magazz\`{u}, Roberto Caruso, Giuseppe Lup\`{o}, Rosalba Saija, Antonino Foti, Pietro Giuseppe Gucciardi, Andrea Mandanici, Onofrio Maria Marag\`{o} \\ and Maria Grazia Donato
\thanks{The authors acknowledge financial contribution from the agreement between Agenzia Spaziale Italiana (ASI) and Istituto Nazionale di Astrofisica (INAF) n.2018-16-HH.0, project “SPACE Tweezers” and from the MSCA innovative training network (ITN) project “ActiveMatter” sponsored by the European Commission (Horizon 2020, Project Number 812780). This work has been also partially funded by European Union (NextGeneration EU), through the MUR-PNRR project SAMOTHRACE (ECS00000022) and PNRR MUR project PE0000023-NQSTI. \textit{(Corresponding authors: Andrea Mandanici and Maria G. Donato)}}
\thanks{S. Marrara, D.Bronte Ciriza, R. Saija and A. Mandanici are with the Dipartimento di Scienze Matematiche e Informatiche, Scienze Fisiche e Scienze della Terra, Università di Messina, I-98166 Messina, Italy and the CNR-IPCF, Istituto per i Processi Chimico-Fisici, I-98158 Messina, Italy (email: sonia.marrara@studenti.unime.it; brontecir@ipcf.cnr.it;  Rosalba.Saija@unime.it; andrea.mandanici@unime.it).}
\thanks{A. Magazz\`{u}, R. Caruso, G. Lup\`{o}, A. Foti, P. G. Gucciardi, O. M. Marag\`{o} and M. G. Donato are with the CNR-IPCF, Istituto per i Processi Chimico-Fisici, I-98158 Messina, Italy (email: magazzu@ipcf.cnr.it; caruso@ipcf.cnr.it; lupo@ipcf.cnr.it; antonino.foti@cnr.it; pietrogiuseppe.gucciardi@cnr.it; onofrio.marago@cnr.it; maria.donato@cnr.it). }
\thanks{S. Marrara and D. Bronte Ciriza contributed equally to this work.}
}

\markboth{Journal of \LaTeX\ Class Files,~Vol.~14, No.~8, August~2021}{S. Marrara \MakeLowercase{\textit{et al.}}: Optical calibration of holographic acoustic tweezers}

\maketitle

\begin{abstract}
Recently, acoustic tweezers based on an array of ultrasonic transducers have been reported taking inspiration from holographic optical tweezers.
In the latter technique, the calibration of the optical trap is an essential procedure to obtain the trap stiffnesses. On the contrary, in the case of acoustic tweezers the calibration of the acoustic forces is seldom carried out. To cover this gap, in this work we adapt the calibration protocols employed in optical tweezers to acoustic tweezers based on arrays of ultrasonic transducers. We measure trap stiffnesses in the mN/m range that are consistent with theoretical estimates obtained by calculations of the acoustic radiation forces based on the Gor'kov potential. This work gives a common framework to the optical and acoustic manipulation communities, paving the way to a consistent calibration of hybrid acousto-optical setups.
\end{abstract}

\begin{IEEEkeywords}
Acoustic tweezers, acoustic levitation, acoustic forces, force calibration, phased arrays, ultrasonics.
\end{IEEEkeywords}

\section{Introduction}
\IEEEPARstart{T}{here} are different ways to trap and manipulate particles: for instance, it is possible to use light \cite{volpe2023roadmap}, but also magnetic \cite{neuman2008single} and electric fields \cite{gerspach2017soft} or a sound wave \cite{brandt2001suspended}. In optical trapping, dielectric particles with size varying from tens of nanometers to few micrometers can be studied; however, its applicability is limited in case of light absorbing materials, due to increased scattering force or particle heating \cite{Jones2015}. Magnetic  and electrostatic  trapping can be used only with magnetic or charged particles, respectively. On the other hand, acoustic trapping requires less stringent conditions on both the size and the type of particle that can be trapped, allowing interesting applications in biomedical research \cite{dholakia2020comparing}, life sciences \cite{thalhammer2016acoustic, ozcelik2018acoustic}, physics of liquids \cite{zang2017acoustic} and soft-matter \cite{meng2019acoustic, ma2020acoustic}.

\begin{figure*}[t]
\centering
\includegraphics[width=0.7\textwidth]{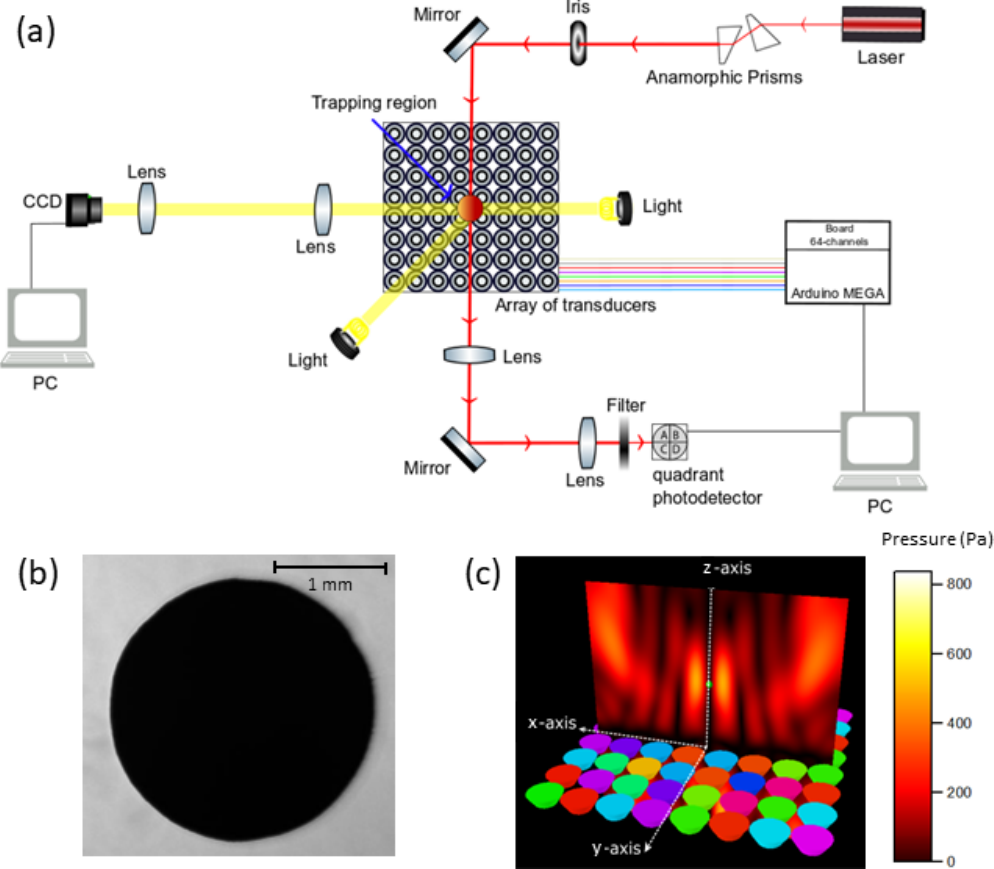}
\caption{a) Scheme of the experimental set-up; b) CCD shadow image of a trapped polystyrene sphere; c) 8x8 transducers array (coloured elements) and map of the produced pressure field (in Pa). The trap reference frame is also represented. The trap is focused at $(x_f,y_f,z_f)$= (0, 0, 2 cm)}
\label{fig1}
\end{figure*}

In acoustic trapping sound waves in the ultrasonic range are used to confine and manipulate millimeter and submillimeter particles in air \cite{koyama2010noncontact, foresti2014acoustophoretic, ospina2022particle} or a fluid \cite{ding2012chip, baresch2016observation, baresch2020acoustic}. The simplest way to obtain particle trapping is the realization of an acoustic standing wave. To this purpose, either an ultrasound source and a reflector \cite{xie2001parametric}, or two sources facing each other \cite{matsui1995translation}, can be used. With these experimental configurations, both liquid and solid particles have been trapped, even with very large densities (Hg and Ir, respectively) \cite{xie2002levitation}. However, these apparatuses have the disadvantage of being expensive and bulky, requiring potentially dangerous high voltages and loosing efficiency after prolonged operation. To overcome these drawbacks, it has been recently reported the possibility of using arrays of commercial ultrasonic transducers both in the standing wave configuration \cite{marzo2017tinylev} and in a single-sided array configuration \cite{marzo2015holographic, Ultraino2018}. In the latter case the shaping of the pressure field is achieved by controlling the phase of the signal emitted by each transducer. This method gets inspiration from  holographic optical tweezers, in which an optical diffractive element (tipically a Spatial Light Modulator) is used to shape an incoming light beam \cite{Jones2015}. With the holographic acoustic tweezers \cite{marzo2015holographic, marzo2019holographic} different acoustic pressure field distributions have been obtained, giving twin traps (two close finger-shaped regions among which the particles are trapped), vortex traps and bottle traps \cite{marzo2015holographic}, which enable the full 3D manipulation of single \cite{marzo2015holographic} or multiple \cite{marzo2019holographic} millimetric particles. The advantage of this configuration with respect to other techniques of acoustic beam shaping \cite{memoli2017metamaterial} is that the acoustic trap is computer-controlled \cite{Ultraino2018, GitHubUltraino}, allowing to change the trap coordinates on demand and in real-time.

In optical tweezers, a routine procedure that enables quantitative force measurements is the calibration of the trap, namely the measurement of its stiffness when a standard size spherical particle is trapped in a medium of known viscosity, and the re-scaling in length units of the particle dynamics by means of appropriate calibration factors \cite{Jones2015}.  Many different protocols have been proposed \cite{Berg2004, tolic2006calibration, Marago2008a, jones2009rotation, Marago2010b, Donato2018}, and a comparison between them can be done \cite{magazzu2015optical}, aiming at individuating the best method to calibrate the particular system under study. On the contrary, in the case of acoustic tweezers, few examples are found in literature \cite{lee2010calibration, li2013simple, lim2016calibration}, treating quite small (tens of $\mu$m) particles in a microfluidic environment. However, in these cases the full spatial localization of the particles is obtained close to a surface, which may affect the estimate of the trap stiffness.

In this work, we tailor two well assessed calibration methodologies developed in optical tweezers to acoustic tweezers setups. After a brief introduction on the methods used to trap and manipulate the particles, we show the results of the calibration of holographic acoustic tweezers  \cite{marzo2015holographic} obtained by both static and dynamic measurements. Finally, we compare these results with acoustic trapping forces calculated on the basis of the Gor'kov potential \cite{bruus2012AF7}.

\section{Methods}
In this section, after some details on the hardware used for the trap calibration (section \ref{expsetup}), we discuss the model used for the acoustic trap (section \ref{model}) and the calculation of the acoustic force (section \ref{force}).

\subsection{Experimental setup}\label{expsetup}
The calibration methodologies employ both video microscopy and a position-sensitive 4-quadrant photodetector to track the fluctuations from the equilibrium position of acoustically trapped particles. The experimental setup is sketched in Fig. \ref{fig1}. The space above an 8$\times $8 array of ultrasonic transducers \cite{marzo2015holographic} is crossed by two perpendicular optical paths. In the first one, a diode laser (785 nm, Thorlabs, DL7140-201S), whose beam is made circular by a couple of anamorphic prisms, is directed towards a levitated millimeter sphere. The shadow of the acoustically trapped particle is imaged, through a mirror (Thorlabs, BB1-E03) and two lenses (Thorlabs, N-BK7, -B coated) in a telescope 3:1 configuration, onto a 4-quadrant photodetector (QPD, RS with home-made analog circuit) that records the axial ($z$ axis) or transversal ($x$ or $y$, depending on the orientation of the trap with respect to the detector) particle position fluctuations. The voltage signal acquired by the QPD is  analyzed by custom-made LabView codes. The second optical path is perpendicular to the first one and consists of two lenses in a 2:1 telescope configuration and a CCD camera (Thorlabs  Zelux CS165MU/M). Two lamps provide front and back illumination. The shadow of the levitated particle is imaged by the CCD, recording movies of the trapped particle with a sampling rate of approximately 100 Hz, after cropping the image. It is worth noting that (see Supplementary Figure 1) whereas in optical tweezers the laser beam allows both particle illumination and trapping (through a high numerical aperture objective), in our case the laser beam (and lamps in the perpendicular detection path) are used only for the particle illumination, aiming at detecting its fluctuations on QPD (or CCD). The trapping of the particle is carried out by the acoustic pressure field produced by the transducers array. Tracking of the particle is obtained with Python-based home-made codes, giving the trajectory of the particle center and the mean particle radius $R$.
The array of transducers has been assembled following the instructions provided in A. Marzo et al.\cite{Ultraino2018} and in the corresponding GitHub repository \cite{GitHubUltraino}. The transducers (Murata MA40S4S), emitting at $f$=40 kHz, are controlled by a suitably programmed Arduino Mega 2560 Rev3 board generating 64 digital periodic signals and a printed circuit mounting 32 TC4427 MOSFET \cite{Ultraino2018} to amplify these signals. The phase of each transducer is changed, according to the chosen shape and focus position of the trap, by means of a Java software \cite{GitHubUltraino}. The circuits are powered by a DC power supply, with voltages ranging from 16 V to a maximum of 20 V.

\subsection{Modeling of the acoustic trap}
\label{model}
To realize the acoustic trap, each $j$-th transducer is modeled as a circular piston, whose far-field acoustic pressure $p_j$ at a target point $A(x, y, z)$ is given by

\begin{equation}
p_{j}=e^{i\phi_{j}}M_{j}
\label{eq.p}
\end{equation}

with $\phi_j$ the initial phase of each transducer and

\begin{equation}
M_{j}=P_{0}V_{pp} \frac{2J_{1}(\kappa \ a \ \mathrm{sin} \theta_{j})}{\kappa \ a\  \mathrm{sin} \theta_{j}} \frac{1}{d_{j}} e^{i \kappa d_{j}}
\label{Eq.M}
\end{equation}

Here, $d_{j}$ is the generic distance of the $j$-th transducer from the $A$ point, $\kappa=\frac{2\pi f}{c_{0}}$ is the wave vector, $f$ is the frequency of the signal emitted by the transducer (40 kHz), $c_0$=346 m/s is the velocity of sound in air, $a$ is the transducer radius, $\theta_{j}$ is the angle between $d_j$ and the normal to the array, $J_1$ is the Bessel function of the first kind, $P_0$=0.17 is a power constant typical of the transducers and $V_{pp}$ is the voltage sent by the power supply.

The initial phase angle $\phi_j$ is chosen in such a way that all the pressure fields $p_j$ are in phase at the same point, the trap focus, at $(x_f,y_f,z_f)$= (0, 0, 2 cm); additional phase contributions need to be taken in account for the generation of twin, vortex or bottle traps \cite{marzo2015holographic}. In the case of a twin trap oriented along the $x$ direction (see Fig. \ref{fig1} c), an additional $\pi$ shift between the initial phases of transducers having $x_j>x_f$ and  $x_j<x_f$, respectively, is used. Once defined all the phases giving a trapping point at the focus, the total pressure field $p$ is calculated by summing the $p_j$ from each transducer, that is, $p=\sum_j p_j$.

\subsection{Calculation of the acoustic force}
\label{force}

In case of spherical particles with radius $R$ smaller than the acoustic wavelength $\lambda$  (in our case, $\lambda=$$c_{0}$/$f$=8.65 mm), a good approximation of the acoustic potential is given by the Gor'kov potential $U$ \cite{bruus2012AF7, jackson2021acoustic}

\begin{equation}
    U = \dfrac{4}{3}\pi R^3 \left[
    b_1 \dfrac{1}{2 c_{0}^{2}\rho_0 }  \langle \left( p \right)^2 \rangle
    - b_2 \dfrac{3}{4} \rho_0 \langle \left(\Vec{u} \right)^2 \rangle
    \right]
    \label{eq:Gorkov_Bruus}
\end{equation}

where $p$ and $\Vec{u}$ are the incident acoustic pressure and the velocity fields at the location of the particle. The factors $b_1$ and $b_2$
are given by
\begin{equation}
    b_1 = 1 - \dfrac{c_{0}^{2}\rho_0}{c_{p}^{2}\rho_p}
\end{equation}
with $c_0$ and  $\rho_0$ the sound velocity and the density in the medium, $c_p$ and $\rho_p$ the sound velocity and the density in the particle, and

\begin{equation}
    b_2 = \dfrac{2\left(\dfrac{\rho_p}{\rho_0} - 1 \right)}{2\dfrac{\rho_p}{\rho_0} + 1}
\end{equation}

In case of harmonic $p$ and $\Vec{u}$ fields and taking in account the relation (see Eq.25 in \cite{jackson2021acoustic})

\begin{equation}
    \rho_0 \dfrac{\partial \Vec{u}}{\partial t} = - \nabla p
\end{equation}

\noindent the Gor'kov potential can be expressed only in terms of the pressure field and its spatial derivatives:

\begin{equation}
    U = K_1 |p|^2 - K_2 \left(
    \bigg| \dfrac{\partial p}{\partial x} \bigg| ^2
    + \bigg| \dfrac{\partial p}{\partial y} \bigg| ^2
    + \bigg| \dfrac{\partial p}{\partial z} \bigg| ^2
    \right)
    \label{eq:Gor_kov}
\end{equation}
with
\begin{equation}
    K_1 = \dfrac{1}{4} V \left(
    \dfrac{1}{\rho_0 c_0^2}
    - \dfrac{1}{\rho_p c_p^2}
    \right)
    \label{K1}
\end{equation}
and
\begin{equation}
    K_2 = \dfrac{3}{4} V \dfrac{1}{\rho_0 \omega^2}
    \dfrac{\rho_p - \rho_0}{\left( 2\rho_p + \rho_0 \right)}
    \label{K2}
\end{equation}
with $V=\frac{4}{3}\pi R^3$ the particle volume and $\omega$ the angular frequency of the pressure wave.

Finally, the acoustic force is obtained as
\begin{equation}
    \Vec{F}=-\nabla U
\end{equation}

\section{Results and Discussion}

An optical trap is characterized by studying the Brownian motion of the trapped particle in the confining optical potential \cite{Jones2015}, which for small displacements from equilibrium can be considered harmonic.

The dynamics of the trapped particle are described by the Langevin equation

\begin{equation}\label{eq:bm:langevinF}
m \frac{d^2}{dt^2}x(t) = -k_x x - \gamma \frac{d}{dt}x(t) + \chi(t) \; .
\end{equation}

where $m$ is the mass of the particle, $k_x$ is the trap spring constant, $\gamma$ is the drag coefficient and $\chi$ is a white noise. For clarity, we have considered only the $x$ direction, but analogous equations can be written for $y$ and $z$ directions, and consequently different trap spring constants $k_y$ and $k_z$ must be considered. At low Reynolds number, that is, when viscous forces on the particles are larger than inertial forces, the inertial term in (\ref{eq:bm:langevinF}) can be neglected, giving the overdamped Langevin equation

\begin{equation}\label{eq:bm:overdamped}
 \frac{dx(t)}{dt} = \displaystyle -\frac{k_x}{\gamma} x(t) + \sqrt{2D} \, W_x(t)
\end{equation}

\noindent where $D=\frac{k_B T}{\gamma}$ is the diffusion coefficient and $W_x(t)$ is the white noise term.

The calibration of the trap is obtained by tracking the trapped particle fluctuations around the trap equilibrium position by means of a CCD camera or by means of a 4-quadrant photodiode. With both devices, the trap spring constants $k_i$ ($i=x,y,z$) can be estimated by fitting the signal Power Spectra (PS) \cite{Jones2015, Berg2004, tolic2006calibration}. The fitting of PS gives also the pixels to meters (in case of the CCD detector) or the Volts to meters (in case of the QPD detector) calibration factors.

Inspired by the optical tweezers calibration protocols, we study the dynamics of a levitated spherical particle by using both video microscopy and the QPD device, and then analyze the obtained trajectories by means of the PS approach. In all measurements, we used a twin trap. The trap reference frame (see Fig. \ref{fig1}c), is chosen in such a way that the $x$ direction is the one connecting the two pressure maxima of the twin trap. The $y$ direction is perpendicular to $x$ and parallel to the array of transducers, while $z$ is the direction perpendicular to the array. As already outlined in the section \ref{expsetup} of the Methods, our CCD and QPD detectors are perpendicular to each other, allowing us to study, for the same twin trap, both the $x$ and the $y$ direction. If we want to choose which direction is observed on each detector, we can rotate the orientation of the trap by 90$ ^\circ$. Fluctuations in the $z$ direction can be observed with both devices.

Calibration measurements consist in the study of the position fluctuations of the particle while it is levitating in the trap \cite{andrade2014experimental}. We study how the particle position fluctuates around the trap equilibrium point when subject to a random perturbation. Moreover, we can also displace in a controlled manner the particle from the trap focus, thanks to the 3D manipulation capability of the setup, and study the damped oscillations of the particle position.  In such a way, we are able to measure both the trap spring constants and provide also an estimation of the air drag coefficient $\gamma$.

For our measurements, we use styrofoam millimeter spheres. These particles are not sold as laboratory standards, and their physical properties are not provided by the producer. However, to calibrate the acoustic trap we need the density of the material. To this aim,
we weighed twenty similar beads on an analytical balance with a sensitivity of 0.1 mg. By averaging on all the particles, we estimate a density of $\rho=36\pm6$ $\mathrm{kg/m^3}$, which is compatible with the values reported by other authors\cite{mihlayanlar2008analysis, zarr2012nist, prasittisopin2022review}.

\subsection{Random perturbations}

\begin{figure}[t]
\centering
\includegraphics[width=0.8\columnwidth]{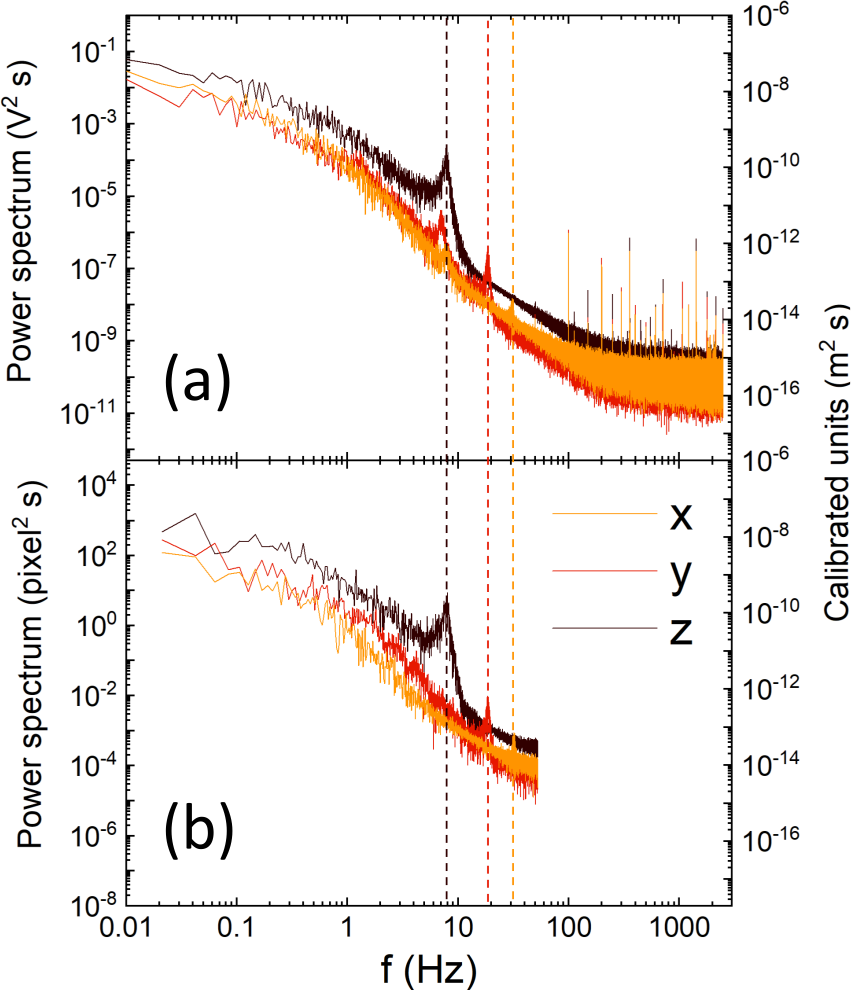}
\caption{Power spectra of spontaneous oscillations of a levitated particle along the trap reference frame $xyz$. a) PS of the signal registered with QPD; b) PS of the signal registered with the video camera. Data are shown both in original units (V$^2$ s in the QPD case and pixel$^2$ s in the CCD case, left axes) and in calibrated units (m$^2$ s, rigth axes). Dashed lines point out correspondence between similar peaks. From data and the harmonic oscillator model, the trap stiffnesses are calculated as $k_x$=16$\pm$2 mN/m, $k_y$=5.2$\pm$0.7 mN/m and $k_z$=0.97$\pm$0.13 mN/m.}
\label{fig2}
\end{figure}

In Fig. \ref{fig2}, the PS calculated by tracking the trajectory of a trapped bead by means of the QPD detector (Fig.\ref{fig2} a) and the CCD camera (Fig.\ref{fig2} b) are shown. PS approach is a very useful tool for the analysis of the tracking signals because a periodic motion of the particle appears as a peak in the spectrum \cite{tolic2006calibration}.

\begin{figure*}
\centering
\includegraphics[width=0.7\textwidth]{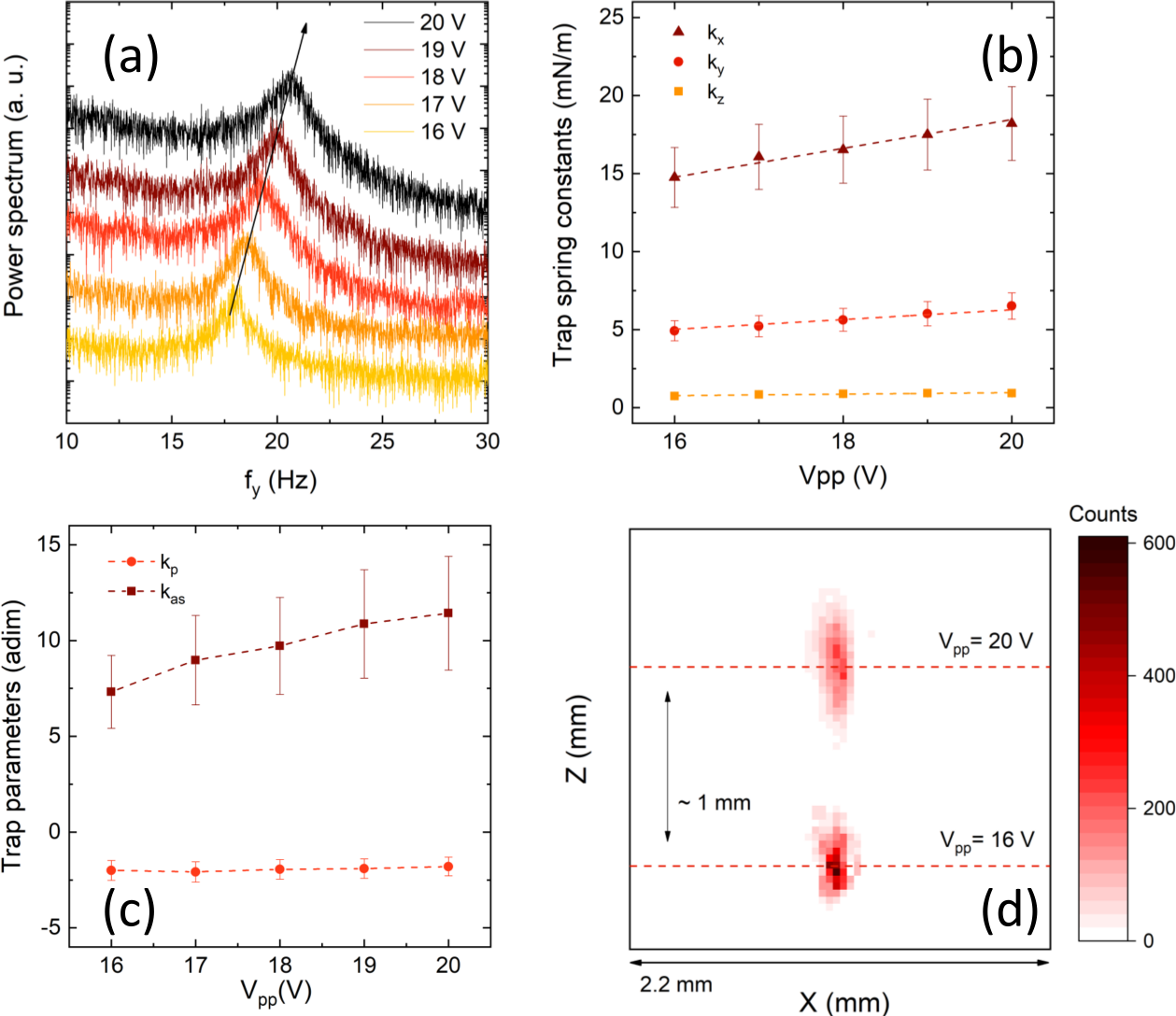}
\caption{a) Power spectra measured at input voltage ranging from 16 V to 20 V for the same trap and particle shown in Fig. \ref{fig2}. The $f_y$ oscillation shifts towards higher values at increasing $V_{pp}$. Similar behaviour is observed for $f_x$ and $f_z$ (not shown). The power spectra have been displaced vertically for clarity. The black arrow is a guide for the eye. b) Trap stiffnesses $k_x$, $k_y$ and $k_z$ as a function of $V_{pp}$. A linear dependence on the input voltage is observed in the voltage range used for measurements. c) Trap parameters $k_p$ and $k_{as}$ as a function of $V_{pp}$. While the transversal asymmetry  pointed out by $k_p$ is approximately constant at increasing $V_{pp}$ (orange circles), the trap aspect ratio, pointed out by $k_{as}$ (red squares), increases at increasing $V_{pp}$. d) Distributions of the trapped particle position. For clarity, they are shown only at the two extreme voltages 16 V and 20 V. The red dashed lines highlight the shift (approx. 1 mm) in the trap equilibrium position at increasing voltage.}
\label{fig3}
\end{figure*}

Here, we show the PS along $x$ (yellow curve), $y$ (red curve), and $z$ (black curve) directions obtained by tracking a levitated particle in a twin trap with focus at $(x_f, y_f, z_f=0,0,2\  \mathrm{cm})$. To obtain the PS along $x$ or $y$ on the same detector, the trap is rotated by changing the signal phases on the transducers.

Peaks at approximately 8 Hz ($f_z$), 18.5 Hz ($f_y$) and 32.4 Hz ($f_x$) are clearly observed on both devices. In the case of the QPD detector, in $x$ and $y$ power spectra a small peak at $f_z$ is also observed, probably due to a small asymmetry of the bead, inducing a tilt on the particle fluctuations and therefore, a cross-talk \cite{Jones2015} between QPD channels. A similar effect is not observed in the signals obtained by the CCD device because the tracking software fits a circle to the particle image and, thus, any particle asymmetry cannot affect the reconstruction of the particle trajectory. Due to the limited sampling rate of CCD (100 Hz) with respect to the QPD (5kHz), the PS calculated with CCD signals are in a shorter frequency range, but retain the most important information, i.e., the frequency peaks $f_x$, $f_y$ and $f_z$. From these peaks and considering a simple harmonic oscillator model, $k_i=4\pi^2f_{i}^{2} m $ $(i=x,y,z)$, the spring constants are obtained as $k_x$=16$\pm$2 mN/m, $k_y$=5.2$\pm$0.7 mN/m and $k_z$=0.97$\pm$0.13 mN/m, as the particle mass is $m$=0.39$\pm$0.05 mg. Finally, the PS obtained with the QPD signals have peaks at frequencies higher than 50 Hz. As these features are observed also without the particle in the trap, or even without laser illumination on the QPD, it is understood that they are periodic spurious signals likely due to electronic noise.

In Fig. \ref{fig3}a the power spectra obtained for particle fluctuations along $Y$ direction are shown at increasing input voltages. As outlined by the black arrow, an increase of the peak frequency $f_y$ with increasing $V_{pp}$ is clearly observed. A similar behaviour has been found also for $f_x$ and $f_z$.
The corresponding trap spring constants $k_x$, $k_y$ and $k_z$ as a function of $V_{pp}$ are shown in Figure \ref{fig3}b. In the voltage range used, a linear dependence of the $k_i$ on $V_{pp}$ is found, even if a quadratic dependence of the trap stiffness on $V_{pp}$ should be observed \cite{bruus2012AF7}, due to the $p^2$ term in the Gor'kov potential. It is worth noting that the voltage range accessible in these measurements is limited by the maximum working voltage of the transducers (20 V) and by the minimum voltage required to stably trap these millimetric styrofoam particles (approximately 15-16 V). Due to this short voltage range, the quadratic dependence of the trap spring constants cannot be observed, whereas an ``effective'' linear dependence is found.

As in optical trapping \cite{RohrbachPRL05, PhysRevLett.100.163903}, the symmetry properties of the the trap can be studied by using parameters connected to the trap spring constants. The parameter $k_p=1- \frac{k_x}{k_y}$ can be used to estimate the symmetry on the $xy$ plane. If the acoustic trap was completely isotropic in this plane, $k_p$ should be zero. We find that this parameter is close to -2, indicating a strong transversal anisotropy of the trap, due to a corresponding anisotropy in the confining potential (see Fig. \ref{fig5}). The trap aspect ratio can be studied by means of the $k_{as}=\frac{k_x+k_y}{2k_z}$ parameter. In our case, we find a quite large value, indicating a prolate shape of the confining potential and of the trapping region in the axial direction, which further become more prolate at increasing $V_{pp}$ (see Fig. \ref{fig3}c and d). Note that at increasing $V_{pp}$ the acoustic force increases while the particle weight remains constant. Thus, the trap equilibrium point changes and, in particular, is located at higher $z$ (Fig. \ref{fig3}d).

\begin{figure}[h]
\centering
\includegraphics[width=0.75\columnwidth]{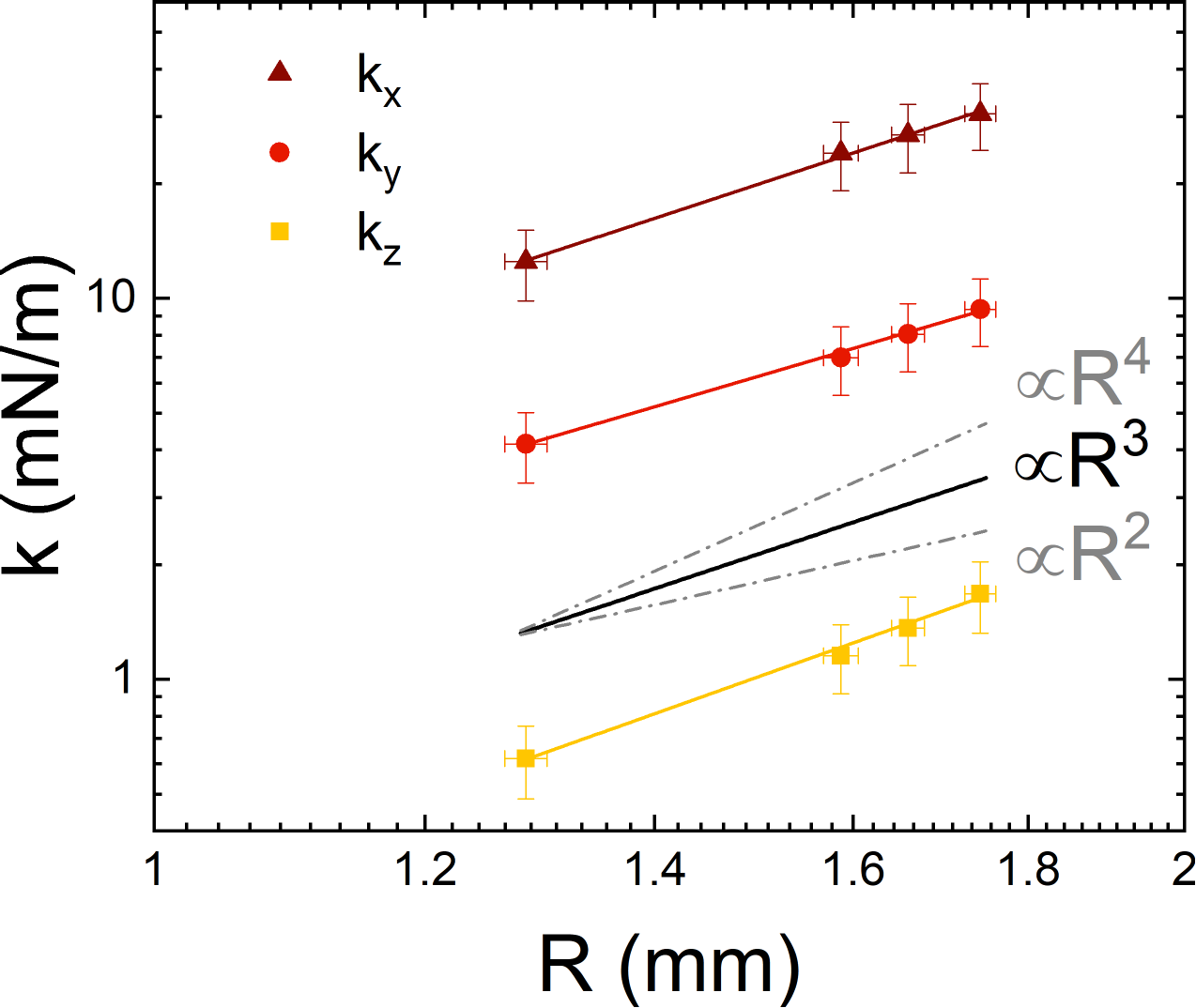}
\caption{Dependence of trap spring constants on particle radius. $k_x$, $k_y$ and $k_z$ are shown  as a function of $R$ for four different particles. To highlight the $k_i$ functional dependence on the radius $R$, different models (black solid line for $R^3$, dashed gray lines for $R^2$ and $R^4$) are plotted as guides to the eye. The $R^3$ dependence of the spring constants is recognized.}
\label{fig4k}
\end{figure}

The Gor'kov potential (\ref{eq:Gor_kov}-\ref{K2}) shows a dependence on the third power of the radius. Thus, a correspondent dependence is also expected on the trap spring constants. To verify this, we trapped four different particles in the same experimental conditions ($V_{pp}$=16 V, trap focus at $x_f, y_f, z_f$= 0, 0, 2 cm) and measured their fluctuations in the trap. The analysis of the PS of the particle fluctuations in the trap allowed the estimation of the trap spring constants, which are shown in Fig. \ref{fig4k} as a function of $R$ in log-log scale. The dependence of the $k_i$ on the third power of the particle radius is recognized.

\subsection{Induced perturbations}
We induced particle position perturbations by periodically changing the position of the trap focus. We shifted the focus by 0.5 mm along $x$, $y$ and $z$ directions at 5 s time intervals and we acquired the consequent damped oscillations of the particle position.

\begin{figure}[h!]
\centering
\includegraphics[width=0.75\columnwidth]{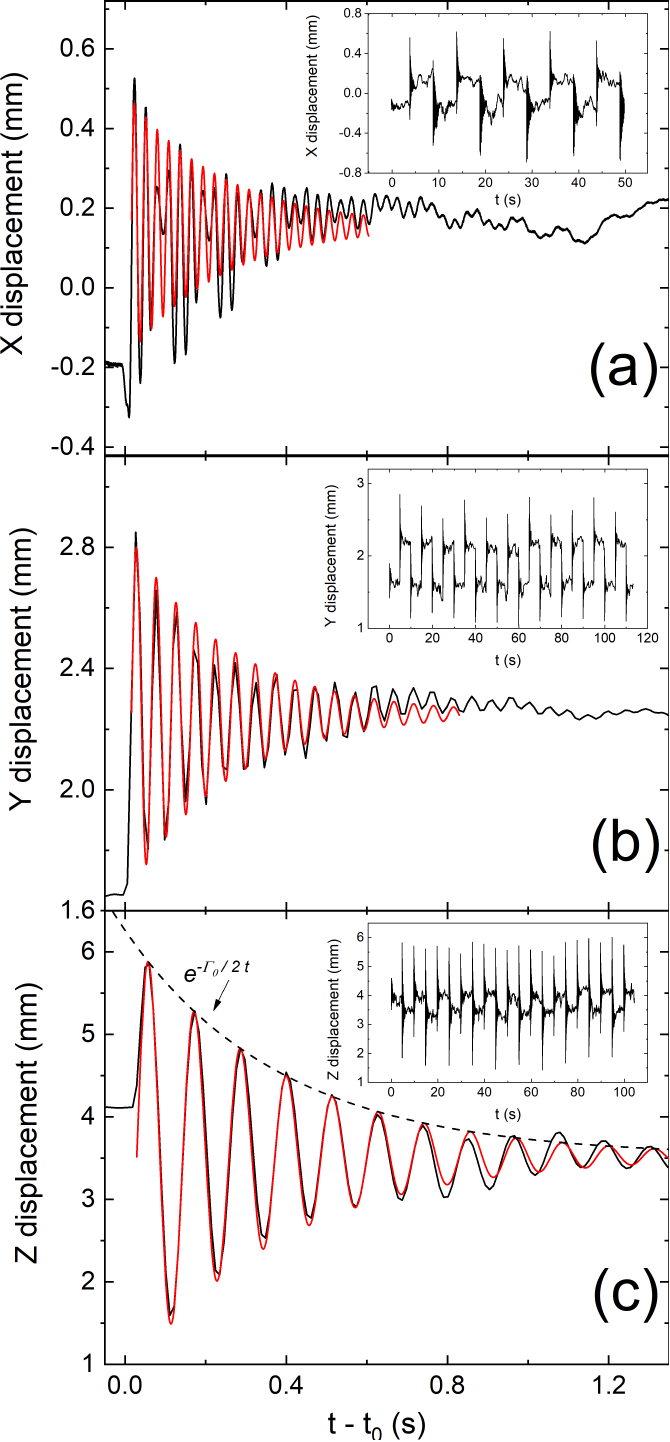}
\caption{Damped harmonic oscillations of an acoustically trapped bead along $x$ (a), $y$ (b) and $z$ (c) directions. The oscillations are induced by periodically changing the particle position along the same directions. The whole traces obtained are shown in the insets. In a), the QPD detector is used, while in b) and c) the data are acquired with the CCD camera. The steps are fitted with a damped harmonic oscillator model (see Eq. \ref{DampHarOsc}) to estimate the oscillator frequency $\omega$ and the damping term $\Gamma_0$. An example of the fitting is given (red curve) for each direction. Note that the time scale is reduced to $t-t_0$ aiming at a comparison between different step traces. Note also that, going from $x$ to $z$,  $\omega$ decreases, pointing out, respectively, a lower $k$. }
\label{fig5}
\end{figure}

In the insets of Figure \ref{fig5} the tracking signals registered on the QPD ($x$ direction) and CCD ($y$ and $z$ directions) are shown.

Note that the faster oscillation is sent on the high acquisition rate detector (the QPD). The signals are characterized by steps, whose initial part (approximately 1 s) is an oscillation with a damped amplitude. After this, the signal becomes similar to what is observed in the random perturbation case. Thus, we decided to focus on the first part of the signal and to fit it with the damped harmonic oscillator model:

\begin{equation}
    y(t) = y_0 + A\  \text{exp}[-\frac{\Gamma_0}{2}  (t-t_0)] \  \text{cos}[\omega (t - t_0) + \varphi]
\label{DampHarOsc}
\end{equation}

\noindent where $A$ is the amplitude, $\Gamma_0=\frac{\gamma}{m}$ the damping coefficient, $\gamma$ the viscous drag coefficient, $m$ the mass of the particle, $\omega$ the angular frequency of the oscillator, $t_0$ the initial time, $\varphi$ a phase constant and $y_0$ a signal offset. Here, $\omega$ is related to the oscillator natural angular frequency $\Omega_0=\sqrt{\frac{k}{m}}$ by the following relation:

\begin{equation}
    \Omega_0^2=\omega^2 +\frac{\Gamma_0^2}{4}
\end{equation}

\begin{figure*}[h]
\centering
\includegraphics[width=0.9\textwidth]{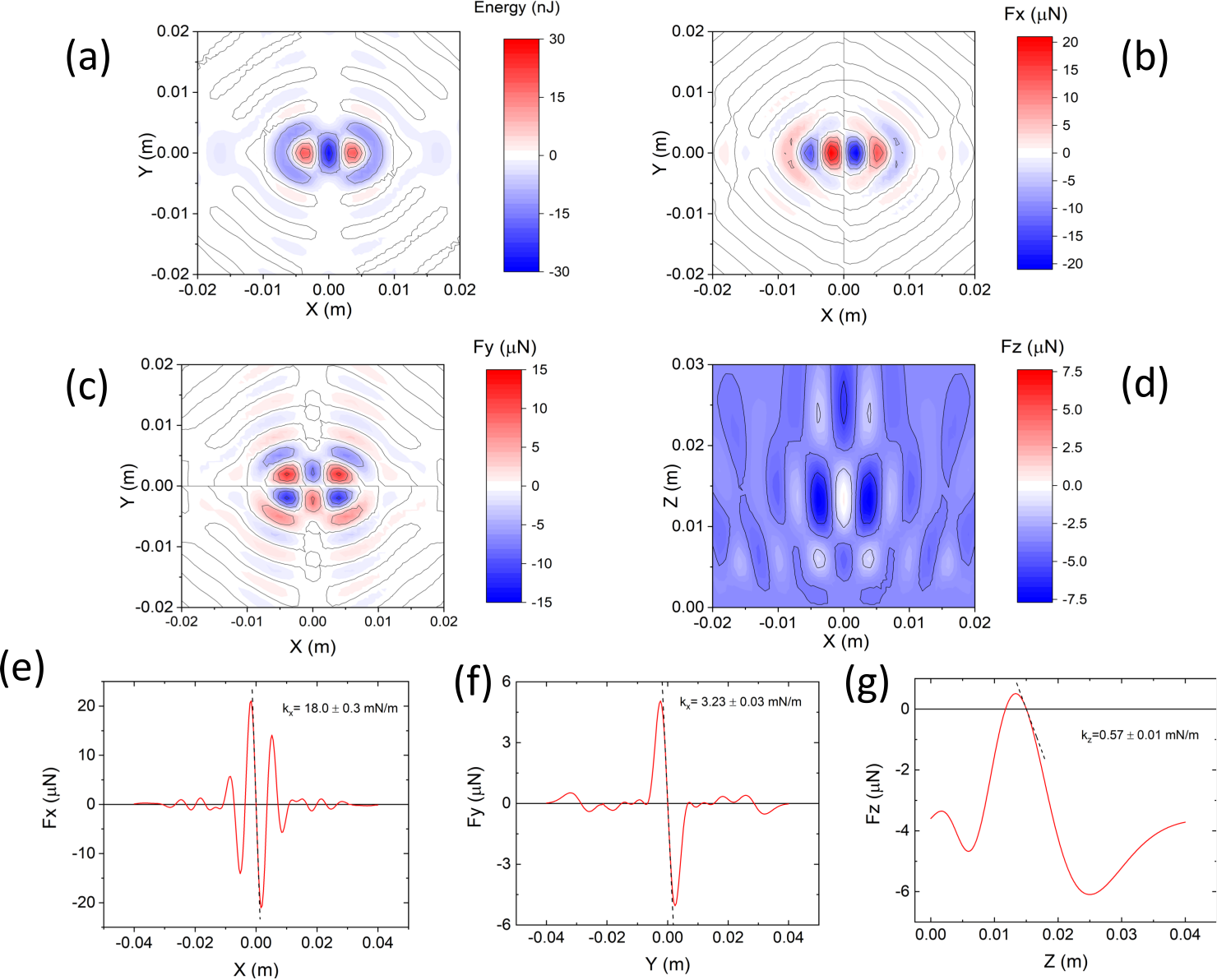}
\caption{Calculation of the a) Gor'kov potential and b-d) correspondent maps of the acoustic force for a $R=$1.36$\pm$0.02 mm bead. The twin trap has been considered in the calculation. Focus has been set at $x_f$=0, $y_f$=0, $z_f$=2 cm above the 8$\times$8 array. e-g) Calculation of the $x$ (e), $y$ (f) and $z$ (g) components of the acoustic force for the same bead. The trap spring constants $k_x$, $k_y$ and $k_z$ have been calculated by a linear fit of the force in the trapping region. Note that in g) also the weight of the particle and the buoyancy have been taken in account. For this reason, the trapping point height is below $z_f$ and located at approximately 1.5 cm.}
\label{fig6}
\end{figure*}

By fitting the particle damped oscillations in $x$, $y$ and $z$ directions we can calculate both the trap spring constants $k_i=\Omega_{0,i}^{2} m$ and also estimate the medium (air) viscous drag coefficient $\gamma$.

In the insets of Fig. \ref{fig5} the whole signals recorded along the three $x$ (a), $y$ (b) and $z$ (c) directions are shown. The step structure of the signals is easily recognized. One of the steps along with its fit (red curve) is also shown for each direction. The frequency of oscillation, and thus the trap spring constant, decreases from $x$ to $z$, in agreement with the results obtained with random perturbations analysis. By averaging on the $\Gamma_0$ and $\omega$ values obtained from the fit of many different steps, the calculated trap spring constants are $k_x=19 \pm 3$ mN/m, $k_y=6 \pm 1$ mN/m, $k_z=1.2 \pm 0.2$ mN/m, which are consistent with results obtained by analyzing only the random perturbations. Moreover, the values of $\Gamma_0$ obtained for each direction can be used to estimate also the average drag coefficient $\gamma=(2.9 \pm 0.8) \cdot 10^{-6}$ N$\cdot$s/m, which has the same order of magnitude and is 6 times larger than the theoretical value $\gamma_0$=0.47$\cdot$10$^{-6}$ N$\cdot$s/m expected for air at 25 $^\circ$C. Note that, by averaging $\gamma$ on the three directions, we model the particle as a perfect sphere, even if we know that some small asymmetries are present. 
This could explain the discrepancy between our result and the expected value for a perfect sphere.

\subsection{Simulation of the acoustic trap.}
To verify the agreement between experimental and theoretical results, we calculated the total pressure field, the Gor'kov potential, the force fields, the components of the acoustic forces along the three spatial directions and the trap spring constants for our 8$\times$8 array of transducers.

For the calculations, we consider a particle with radius $R$=1.36$\pm$0.02 mm, as in our measurements, and we use $\rho_0$=1.18 kg/m$^3$ as the density of the medium, $\rho_p$=36 kg/m$^3$ as the density of the particle material, and $c_p$=900 m/s as the sound velocity in the particle. Finally, we used a voltage supply of $V_{pp}$=18 V.

In Fig. \ref{fig6}a, a $xy$ map of the Gor'kov potential in the $z_f$=2 cm plane is shown. The potential shows an elongated shape which explains the anisotropy found in the transversal trap spring constants. The most stable point is at the focus.

By calculating the gradient of the Gor'kov potential, it is possible to obtain the maps (Fig. \ref{fig6}b-d) of the acoustic force in the $x$, $y$ and $z$ direction. In case of the $z$ direction, we have calculated the total axial force, which takes in account also the gravity and the buoyancy of the particle.  From the maps, it is clear the anisotropic character of the force field, which has a different structure both in the transversal and axial plane.

The origin of the anisotropic trapping in holographic acoustic tweezers is in the anisotropic shape of the pressure field \cite{marzo2015holographic}. Forces along x direction are due to gradients of the pressure field, whereas forces along y and z direction are due to gradients of the velocity field \cite{marzo2015holographic}, with velocity itself obtained as the gradient of the pressure field (6). Moreover, along z also gravity and buoyancy control the trap equilibrium position. Thus, anisotropy in the trapping forces along the three spatial directions is expected and confirmed by the anisotropy of the trapping potential (see Fig. \ref{fig6} a).

To individuate the trap equilibrium point and to calculate the trap spring constants, we can plot the $F_x$, $F_y$ and $F_z$ components of the acoustic force along $x$, $y$ and $z$ directions, respectively. The point where the force vanishes with a negative slope is a stable equilibrium point and corresponds to the trap position. By linearly fitting the curves at the trap position we can determine the trap spring constants. In Figs. \ref{fig6}(e)-\ref{fig6}(g) the forces (red curves), the linear fits (dashed black lines) and the trap spring constants calculated (insets) are shown. The values obtained agree fairly well with the measured values, confirming that the Gor'kov potential approximation can be used to model the acoustic trapping of millimetric particles in air with this setup.

\section{Conclusions}
In this work, we have presented a calibration procedure for an acoustic tweezers set-up based on a flat array of transducers. The procedure is inspired by protocols used for optical tweezers. The trap spring constants and the medium drag coefficient have been measured by using both random perturbations and controlled displacements from the equilibrium position. We have shown that the Gor'kov potential and the acoustic force calculated from it are a good approximation to acoustic forces that come into play in real setups. We think that these protocols, sharing the same tools used for optical tweezers, may bridge the gap between the two optical and acoustic trapping communities, paving the way to the realization of hybrid setups where both light and sound can be used to trap particles in air.

\bibliographystyle{IEEEtran}

\begin{thebibliography}{10}
\providecommand{\url}[1]{#1} \csname url@samestyle\endcsname \providecommand{\newblock}{\relax}
\providecommand{\bibinfo}[2]{#2}
\providecommand{\BIBentrySTDinterwordspacing}{\spaceskip=0pt\relax}
\providecommand{\BIBentryALTinterwordstretchfactor}{4}
\providecommand{\BIBentryALTinterwordspacing}{\spaceskip=\fontdimen2\font plus
\BIBentryALTinterwordstretchfactor\fontdimen3\font minus
  \fontdimen4\font\relax}
\providecommand{\BIBforeignlanguage}[2]{{%
\expandafter\ifx\csname l@#1\endcsname\relax
\typeout{** WARNING: IEEEtran.bst: No hyphenation pattern has been}%
\typeout{** loaded for the language `#1'. Using the pattern for}%
\typeout{** the default language instead.}%
\else \language=\csname l@#1\endcsname \fi #2}} \providecommand{\BIBdecl}{\relax} \BIBdecl

\bibitem{volpe2023roadmap}
G.~Volpe, O.~M. Marago, H.~Rubinsztein-Dunlop, G.~Pesce, A.~Stilgoe, G.~Volpe,
  G.~Tkachenko, V.~G. Truong, S.~Nic~Chormaic, F.~Kalantarifard \emph{et~al.},
  ``Roadmap for optical tweezers 2023,'' \emph{Journal of Physics: Photonics},
  2023.

\bibitem{neuman2008single}
K.~C. Neuman and A.~Nagy, ``Single-molecule force spectroscopy: optical
  tweezers, magnetic tweezers and atomic force microscopy,'' \emph{Nature
  Methods}, vol.~5, no.~6, pp. 491--505, 2008.

\bibitem{gerspach2017soft}
M.~A. Gerspach, N.~Mojarad, D.~Sharma, T.~Pfohl, and Y.~Ekinci, ``Soft
  electrostatic trapping in nanofluidics,'' \emph{Microsystems \&
  Nanoengineering}, vol.~3, no.~1, pp. 1--10, 2017.

\bibitem{brandt2001suspended}
E.~Brandt, ``Suspended by sound,'' \emph{Nature}, vol. 413, no. 6855, pp.
  474--475, 2001.

\bibitem{Jones2015}
P.~H. Jones, O.~M. Marag\`{o}, and G.~Volpe, \emph{Optical tweezers: Principles
  and applications}.\hskip 1em plus 0.5em minus 0.4em\relax Cambridge:
  Cambridge University Press, 2015.

\bibitem{dholakia2020comparing}
K.~Dholakia, B.~W. Drinkwater, and M.~Ritsch-Marte, ``Comparing acoustic and
  optical forces for biomedical research,'' \emph{Nature Reviews Physics},
  vol.~2, no.~9, pp. 480--491, 2020.

\bibitem{thalhammer2016acoustic}
G.~Thalhammer, C.~McDougall, M.~P. MacDonald, and M.~Ritsch-Marte, ``Acoustic
  force mapping in a hybrid acoustic-optical micromanipulation device
  supporting high resolution optical imaging,'' \emph{Lab on a Chip}, vol.~16,
  no.~8, pp. 1523--1532, 2016.

\bibitem{ozcelik2018acoustic}
A.~Ozcelik, J.~Rufo, F.~Guo, Y.~Gu, P.~Li, J.~Lata, and T.~J. Huang, ``Acoustic
  tweezers for the life sciences,'' \emph{Nature Methods}, vol.~15, no.~12, pp.
  1021--1028, 2018.

\bibitem{zang2017acoustic}
D.~Zang, Y.~Yu, Z.~Chen, X.~Li, H.~Wu, and X.~Geng, ``Acoustic levitation of
  liquid drops: Dynamics, manipulation and phase transitions,'' \emph{Advances
  in colloid and interface science}, vol. 243, pp. 77--85, 2017.

\bibitem{meng2019acoustic}
L.~Meng, F.~Cai, F.~Li, W.~Zhou, L.~Niu, and H.~Zheng, ``Acoustic tweezers,''
  \emph{Journal of Physics D: Applied Physics}, vol.~52, no.~27, p. 273001,
  2019.

\bibitem{ma2020acoustic}
Z.~Ma, A.~W. Holle, K.~Melde, T.~Qiu, K.~Poeppel, V.~M. Kadiri, and P.~Fischer,
  ``Acoustic holographic cell patterning in a biocompatible hydrogel,''
  \emph{Advanced Materials}, vol.~32, no.~4, p. 1904181, 2020.

\bibitem{koyama2010noncontact}
D.~Koyama and K.~Nakamura, ``Noncontact ultrasonic transportation of small
  objects over long distances in air using a bending vibrator and a
  reflector,'' \emph{IEEE transactions on Ultrasonics, Ferroelectrics, and
  Frequency control}, vol.~57, no.~5, pp. 1152--1159, 2010.

\bibitem{foresti2014acoustophoretic}
D.~Foresti and D.~Poulikakos, ``Acoustophoretic contactless elevation, orbital
  transport and spinning of matter in air,'' \emph{Physical Review Letters},
  vol. 112, no.~2, p. 024301, 2014.

\bibitem{ospina2022particle}
J.~F.~P. Ospina, V.~Contreras, J.~Estrada-Morales, D.~Baresch, J.~L. Ealo, and
  K.~Volke-Sep{\'u}lveda, ``Particle-size effect in airborne standing-wave
  acoustic levitation: Trapping particles at pressure antinodes,''
  \emph{Physical Review Applied}, vol.~18, no.~3, p. 034026, 2022.

\bibitem{ding2012chip}
X.~Ding, S.-C.~S. Lin, B.~Kiraly, H.~Yue, S.~Li, I.-K. Chiang, J.~Shi, S.~J.
  Benkovic, and T.~J. Huang, ``On-chip manipulation of single microparticles,
  cells, and organisms using surface acoustic waves,'' \emph{Proceedings of the
  National Academy of Sciences}, vol. 109, no.~28, pp. 11\,105--11\,109, 2012.

\bibitem{baresch2016observation}
D.~Baresch, J.-L. Thomas, and R.~Marchiano, ``Observation of a single-beam
  gradient force acoustical trap for elastic particles: acoustical tweezers,''
  \emph{Physical Review Letters}, vol. 116, no.~2, p. 024301, 2016.

\bibitem{baresch2020acoustic}
D.~Baresch and V.~Garbin, ``Acoustic trapping of microbubbles in complex
  environments and controlled payload release,'' \emph{Proceedings of the
  National Academy of Sciences}, vol. 117, no.~27, pp. 15\,490--15\,496, 2020.

\bibitem{xie2001parametric}
W.~Xie and B.~Wei, ``Parametric study of single-axis acoustic levitation,''
  \emph{Applied Physics Letters}, vol.~79, no.~6, pp. 881--883, 2001.

\bibitem{matsui1995translation}
T.~Matsui, E.~Ohdaira, N.~Masuzawa, and M.~I.~M. Ide, ``Translation of an
  object using phase-controlled sound sources in acoustic levitation,''
  \emph{Japanese Journal of Applied Physics}, vol.~34, no.~5S, p. 2771, 1995.

\bibitem{xie2002levitation}
W.~Xie, C.~Cao, Y.~L{\"u}, and B.~Wei, ``Levitation of iridium and liquid
  mercury by ultrasound,'' \emph{Physical Review Letters}, vol.~89, no.~10, p.
  104304, 2002.

\bibitem{marzo2017tinylev}
A.~Marzo, A.~Barnes, and B.~W. Drinkwater, ``Tiny{L}ev: A multi-emitter
  single-axis acoustic levitator,'' \emph{Review of Scientific Instruments},
  vol.~88, no.~8, p. 085105, 2017.

\bibitem{marzo2015holographic}
A.~Marzo, S.~A. Seah, B.~W. Drinkwater, D.~R. Sahoo, B.~Long, and
  S.~Subramanian, ``Holographic acoustic elements for manipulation of levitated
  objects,'' \emph{Nature Communications}, vol.~6, no.~1, pp. 1--7, 2015.

\bibitem{Ultraino2018}
A.~Marzo, T.~Corkett, and B.~W. Drinkwater, ``Ultraino: An open phased-array
  system for narrowband airborne ultrasound transmission,'' \emph{IEEE
  Transactions on Ultrasonics, Ferroelectrics, and Frequency Control}, vol.~65,
  no.~1, pp. 102--111, 2018.

\bibitem{marzo2019holographic}
A.~Marzo and B.~W. Drinkwater, ``Holographic acoustic tweezers,''
  \emph{Proceedings of the National Academy of Sciences}, vol. 116, no.~1, pp.
  84--89, 2019.

\bibitem{memoli2017metamaterial}
G.~Memoli, M.~Caleap, M.~Asakawa, D.~R. Sahoo, B.~W. Drinkwater, and
  S.~Subramanian, ``Metamaterial bricks and quantization of meta-surfaces,''
  \emph{Nature Communications}, vol.~8, no.~1, pp. 1--8, 2017.

\bibitem{GitHubUltraino}
\BIBentryALTinterwordspacing A.~Marzo, ``Ultraino,'' 2017, (Last viewed 2022-08-12). [Online].
Available:
  \url{https://github.com/asiermarzo/Ultraino}
\BIBentrySTDinterwordspacing

\bibitem{Berg2004}
K.~Berg-Sorensen and H.~Flyvbjerg, ``Power spectrum analysis for optical
  tweezers,'' \emph{Review of Scientific Instruments}, vol.~75, p. 594, 2004.

\bibitem{tolic2006calibration}
S.~F. Toli{\'c}-N{\o}rrelykke, E.~Sch{\"a}ffer, J.~Howard, F.~S. Pavone,
  F.~J{\"u}licher, and H.~Flyvbjerg, ``Calibration of optical tweezers with
  positional detection in the back focal plane,'' \emph{Review of Scientific
  Instruments}, vol.~77, no.~10, p. 103101, 2006.

\bibitem{Marago2008a}
O.~M. Marag\`{o}, P.~H. Jones, F.~Bonaccorso, V.~Scardaci, P.~G. Gucciardi,
  A.~G. Rozhin, and A.~C. Ferrari, ``Femtonewton force sensing with optically
  trapped nanotubes,'' \emph{Nano Letters}, vol.~8, pp. 3211--3216, 2008.

\bibitem{jones2009rotation}
P.~Jones, F.~Palmisano, F.~Bonaccorso, P.~Gucciardi, G.~Calogero, A.~Ferrari,
  and O.~Marago, ``Rotation detection in light-driven nanorotors,'' \emph{ACS
  Nano}, vol.~3, no.~10, pp. 3077--3084, 2009.

\bibitem{Marago2010b}
O.~M. Marag\`{o}, F.~Bonaccorso, R.~Saija, G.~Privitera, P.~G. Gucciardi, M.~A.
  Iat\`{\i}, G.~Calogero, P.~H. Jones, F.~Borghese, P.~Denti, V.~Nicolosi, and
  A.~C. Ferrari, ``Brownian motion of graphene,'' \emph{ACS Nano}, vol.~4, pp.
  7515--7523, 2010.

\bibitem{Donato2018}
\BIBentryALTinterwordspacing M.~G. Donato, E.~Messina, A.~Foti, T.~J. Smart, P.~H. Jones, M.~A.
Iat\`{\i}, R.~Saija, P.~G. Gucciardi, and O.~M. Marag\`{o}, ``Optical trapping and optical force
positioning of two-dimensional materials,'' \emph{Nanoscale}, vol.~10, pp. 1245--1255, 2018.
[Online]. Available: \url{http://dx.doi.org/10.1039/C7NR06465A} \BIBentrySTDinterwordspacing

\bibitem{magazzu2015optical}
A.~Magazz{\'u}, D.~Spadaro, M.~G. Donato, R.~Sayed, E.~Messina, C.~D'Andrea,
  A.~Foti, B.~Fazio, M.~A. Iat{\'\i}, A.~Irrera \emph{et~al.}, ``Optical
  tweezers: a non-destructive tool for soft and biomaterial investigations,''
  \emph{Rendiconti Lincei}, vol.~26, no.~2, pp. 203--218, 2015.

\bibitem{lee2010calibration}
J.~Lee, C.~Lee, and K.~K. Shung, ``Calibration of sound forces in acoustic
  traps,'' \emph{IEEE transactions on Ultrasonics, Ferroelectrics, and
  Frequency control}, vol.~57, no.~10, pp. 2305--2310, 2010.

\bibitem{li2013simple}
Y.~Li, C.~Lee, K.~Ho~Lam, and K.~Kirk~Shung, ``A simple method for evaluating
  the trapping performance of acoustic tweezers,'' \emph{Applied Physics
  Letters}, vol. 102, no.~8, p. 084102, 2013.

\bibitem{lim2016calibration}
H.~G. Lim, Y.~Li, M.-Y. Lin, C.~Yoon, C.~Lee, H.~Jung, R.~H. Chow, and K.~K.
  Shung, ``Calibration of trapping force on cell-size objects from
  ultrahigh-frequency single-beam acoustic tweezer,'' \emph{IEEE Transactions
  on Ultrasonics, Ferroelectrics, and Frequency Control}, vol.~63, no.~11, pp.
  1988--1995, 2016.

\bibitem{bruus2012AF7}
H.~Bruus, ``Acoustofluidics 7: The acoustic radiation force on small
  particles,'' \emph{Lab on a Chip}, vol.~12, no.~6, pp. 1014--1021, 2012.

\bibitem{jackson2021acoustic}
D.~P. Jackson and M.-H. Chang, ``Acoustic levitation and the acoustic radiation
  force,'' \emph{American Journal of Physics}, vol.~89, no.~4, pp. 383--392,
  2021.

\bibitem{andrade2014experimental}
M.~A. Andrade, N.~P{\'e}rez, and J.~C. Adamowski, ``Experimental study of the
  oscillation of spheres in an acoustic levitator,'' \emph{The Journal of the
  Acoustical Society of America}, vol. 136, no.~4, pp. 1518--1529, 2014.

\bibitem{mihlayanlar2008analysis}
E.~M{\i}hlayanlar, {\c{S}}.~Dilma{\c{c}}, and A.~G{\"u}ner, ``Analysis of the
  effect of production process parameters and density of expanded polystyrene
  insulation boards on mechanical properties and thermal conductivity,''
  \emph{Materials \& Design}, vol.~29, no.~2, pp. 344--352, 2008.

\bibitem{zarr2012nist}
R.~Zarr and A.~Pintar, ``Nist special publication 260-175,'' \emph{Standard
  Reference Materials: SRM}, vol. 1453, 2012.

\bibitem{prasittisopin2022review}
L.~Prasittisopin, P.~Termkhajornkit, and Y.~H. Kim, ``Review of concrete with
  expanded polystyrene (eps): Performance and environmental aspects,''
  \emph{Journal of Cleaner Production}, p. 132919, 2022.

\bibitem{RohrbachPRL05}
A.~Rohrbach, ``{Stiffness of Optical Traps: Quantitative Agreement between
  Experiment and Electromagnetic Theory},'' \emph{Physical Review Letters},
  vol.~95, p. 168102, 2005.

\bibitem{PhysRevLett.100.163903}
F.~Borghese, P.~Denti, R.~Saija, M.~A. Iat\`{\i}, and O.~M. Marag\`o,
  ``Radiation torque and force on optically trapped linear nanostructures,''
  \emph{Phys. Rev. Lett.}, vol. 100, p. 163903, Apr 2008.

\end{thebibliography}


\end{document}